# Electromagnetic Waves in Hot and Dense Media

**Samina Masood** [1,*]

[1] Affiliation 1; masood@uhcl.edu
[*] Correspondence: masood@uhcl.edu; Tel.: (+1281-283-3781)

**Abstract:** It is known that the finite temperature and density (FTD) corrections increase the electron mass, charge and modify the properties of the emitted radiation. All the signals, travelling through the astronomical bodies, carry over the information of their origin and bring minor details about the structure and composition of the source. It has been noticed that temperatures of the early universe add physically measureable mass to electron and large chemical potential lead to an increase in mass as well. However, the QED coupling is slowly increased with temperature but is decreased with increasing chemical potential in superdense systems. Due to the strong relationship between the properties of electromagnetic signals and composition of the material of the propagating objects, we propose to incorporate the modified signal properties to analyze the astronomical data and its interpretation. Existence of different phases of stellar matter at the quantum scale indicates frequent phase transitions for rapidly rotating superdense materials. This information is related to the fact that the allowed range of temperature and chemical potential for a physical system is determined by the signal properties and properties changes with the material as well..

**Keywords:** Propagation of light; Hot and dense media; Astrophysical systems, Early universe, Compact stellar objects, Nanomaterials; Observational data analysis.

## 1. Introduction

Properties of electromagnetic waves are modified when they propagate through extremely hot and dense media with strong magnetic fields. Emitted or scattered radiation from an object is considered to be a reliable source of information for distant objects like stars and galaxies and may determine the behavior and composition of the early universe. The electromagnetic radiations (em-radiation) provides indirect knowledge about the point of its emergence. The effect of statistical properties of a medium on the measureable macroscopic properties of signals such as wavelength, frequency, angle of scattering, refractive index and the energy of the wave is used to determine the structure and dynamics of the system itself.

The complete range of energies and frequencies of electromagnetic spectrum, varying from radio waves to gamma rays, provide the best tool to collect data to study different aspects of astronomical bodies. Light is transverse in nature and reflection, refraction and transmission are the basic properties of electromagnetic waves. An em-signal, passing through a material undergoes deflection based on its index of refraction and scattering probability, associated with the properties and the distribution of the material in space. Propagation, deflection and scattering of light in a medium depend on the properties of a medium. The electromagnetic properties, chemical composition and structure, shape and size of the material objects have a strong impact on the deflection of light in a material and the speed of signal propagation inside the material. On the other hand, scattering of light is considered to be a very powerful technique to find the structure of regular materials in laboratory setup. This interaction of low energy electromagnetic waves does not modify the chemical structure and helps to determine the physical and chemical properties very well. Similar information sources are used to study the stellar composition and structure but astronomical bodies contain much different material than the laboratory materials. Active stellar cores, in particular, is composed of extremely high-energy nuclear matter and outside the core, they have interacting high-energy fluids.

Superdense stellar objects have subatomic nuclear matter in the active portions of the stellar interior. Nuclear fusion processes are required for the burning of star and we do not discuss that type of stellar activity to study QED. However, this material may be a source of electrons through nuclear processes and electrons can propagate through the nuclear matter. The temperatures and densities of such systems are so high that this nuclear matter can exist in the form of quark-gluon plasma or superfluids [1-2], instead of bound states of atoms or molecules.

In the presence of subatomic or nuclear matter, several phases of stellar matter can simultaneously exist locally, occupying tiny regions on quantum scale. Highly dynamic rotating systems of superdense stars like supernovae and neutron stars can have local phase transitions in nanosized regions. For the detailed study of such stars, interaction of radiation with matter has to be incorporated and the properties of matter and radiation are modified with the variation in local statistical conditions. The interaction of extremely high energy radiations with stellar matter occurs at extremely high energies and may go deeper than chemical to fundamental particles.

We reinvestigate the interaction of radiation with matter in the early universe and in stellar interiors, especially the superdense stellar cores [3-5] and develop techniques to incorporate medium effects to interpret the observational data. Fundamental properties of electromagnetic signals and the effect of the propagating medium on these properties, play pivotal role in the analysis of astronomical data. A major difference occurs due to the unusual interaction of radiation with matter at the individual particle level and its impact on the collective behavior of the fluid via electromagnetic properties of the super-hot and super-dense media is studied.

The longitudinal component of electromagnetic waves is zero in free space due to its transverse nature. Chemical properties are measureable in the laboratory frame at macroscopic level but at nanoscale and beyond, individual particle interactions are not ignorable. For astrophysical objects, we have limitations in collecting information about the composition and structure of astrophysical materials. Modification in the properties of electromagnetic waves in the extremely hot and dense materials of the stellar objects should be incorporated to determine the accurate structure and composition of stars. It has been noticed that the properties of light undergo drastic changes at astrophysical temperatures and densities [6-14].

In this paper, we study the propagation of light in hot and dense medium in the range of temperatures above one billion kelvin and chemical potentials of the order of million electron volts, which is above $10^{10}$ K. In such a highly dense quantum statistical medium, light develops a longitudinal component due to its direct interaction with the medium through virtual electrons (during vacuum polarization) and the light disperses differently in different directions in space, as described by the modified form of vacuum polarization tensor in a medium [8] and discussed in the next section. It means that it is not a uniformly distributed sphere in free space, it undergoes modifications based on the local conditions of temperatures and densities in the surroundings.

In the next sections, we give the mathematical scheme of calculations and then show that how light develops longitudinal component when it interacts with an extremely hot and dense medium [8] of astrophysical systems. We make use of the previously calculated mathematical expressions [12-18] for different phases of matter and evaluate them for the relevant astronomical objects to investigate the details about their composition and structure. Section 3 is devoted to the calculations of the relevant QED parameters of extremely hot and dense systems and proposed to adjust the relevant observational data. The mechanism of propagation of light in a hot and dense medium are discussed in Section 5. We have presented all the results of QED calculations in Section 6 and studied the propagation of light in Section 6 and then the results and discussions are included in section 7 followed by the conclusion of results afterwards.

1. **Mathematical Framework**

Effective parameters of QED in the background of hot and dense media are evaluated using the renormalization scheme of QED in the form of the renormalization constants of the theory [4]. In an ultra-relativistic electron-photon system, named as QED system, we evaluate these constants in different ranges of temperature and chemical potential and compare them to the electron mass. Feynman rules of QED are used by replacing Maxwell-Boltzmann distribution by the Bose-Einstein distribution for bosons and Fermi-Dirac distribution for fermions to incorporate the temperature and density dependence of quantum statistical background for interacting fluid.

QED lagrangian, in the usual notation of field theory, with the renormalized values of all QED parameters is expressed as:

$$\mathcal{L} = -\frac{1}{4}F^{\mu\nu}F_{\mu\nu} - \bar{\psi}[\gamma_\mu \partial^\mu + m]\psi - ieA_\mu\bar{\psi}\gamma^\mu\psi - \frac{1}{4}(Z_3 - 1)F^{\mu\nu}F_{\mu\nu} - (Z_2 - 1)\bar{\psi}[\gamma_\mu \partial^\mu + m]\psi + Z_2\delta m\bar{\psi}\psi - ie(Z_2 - 1)A_\mu\bar{\psi}\gamma^\mu\psi$$

and, $F^{\mu\nu} = \partial^\mu A_\mu - \partial^\nu A_\nu$, with $E_i = cF_{0i}$, and $B_i = \frac{1}{2}\epsilon_{ijk}F^{jk}$,

where $\delta m, Z_2$ and $Z_3$ correspond to the renormalization constants of electron mass, charge and wavefunction of electron and are proved to depend on the temperature and chemical potential of electron in the early universe and superdense stars.

Quantum statistical field theory is a gateway to calculate the finite temperature and density (FTD) corrections using the relativistic distribution functions of Bose-Einstein distribution and Fermi-Dirac distribution functions, and incorporate the possibility of interactions with bosons and fermions of the medium, depending on the spin of the particles, respectively. The distribution functions are incorporated in the propagator due to the probability of hot particles with matching attributes from the background. These background contributions, occur virtually through the perturbative processes and we can rewrite the propagators in terms of the relevant distribution functions based on their spin statistics. The photon propagator in Landau gauge is given as:

$$D_\beta^{\mu\nu}(q) = -g^{\mu\nu}\left[\frac{i}{q^2+i\epsilon} + 2\pi\delta(q^2)n_B(E_q)\right] = -g^{\mu\nu}\left[\frac{i}{q^2+i\epsilon} + \Gamma_B(K,\beta)\right] \quad (1a)$$

and the fermion propagator is written as

$$S_F(p,\beta,\mu) = (\gamma^\mu p_\mu - m)\left[\frac{i}{p^2-m^2} - \Gamma_F(p,\beta,\mu)\right] \quad (1b)$$

Where $\beta = 1/T$ throughout this paper. Of course, Boltzmann constant is set to unity in natural units.

The boson distribution function for photons in the background with momentum k is given as:

$$n_B(k) = \frac{1}{e^{\beta k}-1} = \sum_n (-1)^{n+1} e^{-n\beta k} \qquad (2a)$$

and the corresponding contribution to the photon propagator is given as

$$\Gamma_B(K,\beta) = 2\pi i \delta(K^2) n_B(k) \qquad (2b)$$

And the fermion distribution function helps to incorporate interaction with the electrons in the medium and is given as:

$$n_F(p, \pm \mu) = \frac{1}{e^{\beta(E,\pm\mu)}+1} = \sum_n (-1)^n e^{-n\beta(E,\pm\mu)} \qquad (3a)$$

Describing $\Gamma_F(p, \beta, \mu)$ as [11]:

$$\Gamma_F(p, \beta, \mu) = 2\pi i \delta(p^2 - m^2)[\theta(p_0) n_F(p, \mu) + \theta(-p_0) n_F(p, -\mu)] \qquad (3b)$$

$\theta(p_0)$ is a Heaviside function or step function. The chemical potential $\mu$ is the electron chemical potential and $-\mu$ corresponds to the positron chemical potential. First term in eq. (3b) gives the electron background contribution and the second term gives the positron contribution. Only one term of eq. (3) contributes from the background at a time.

In the perturbative quantum electrodynamics (QED), the vacuum polarization tensor is expressed in the usual notation of QED as:

$$\pi_{\mu\nu}(K, T, \mu) = ie^2 \int \frac{d^4p}{(4\pi)^4} \text{Tr}\left\{\gamma_\mu(\gamma_\alpha p^\alpha + \gamma_\alpha K^\alpha + m)\gamma_\nu(\gamma_\alpha p^\alpha + m)\right\}\left[\frac{1}{(p+K)^2 - m^2} + \Gamma_F(p+K, \mu)\right]\left[\frac{1}{p^2 - m^2} + \Gamma_F(p, \mu)\right] \qquad (4)$$

Background effects are included through the propagator term $\Gamma_F(p, \mu)$ and thermal and chemical potential contributions is calculated using Eqns. (2) and (3).

When electromagnetic waves propagate through an interacting medium, they pick up the parameters of the system such as the modified value of background electron mass, wavefunction and charge. In the next section, we have included an overview of the QED parameters in a hot and dense media of astrophysical importance. Then the calculation of a few other parameters of the interacting fluid is presented in the next section in the relevant ranges of statistical media.

2. **Finite Temperature and Density Dependent QED Parameters**

QED fluids are electromagnetically interacting fluids at extremely high temperatures and densities, usually relevant for the stellar interior and the early universe. They are highly compressed materials at extremely high temperatures of the order of billion K or more and densities are above $10^{14}$ kg/m³. These high densities made it possible to generate highly magnetized materials of densities of the order of billion gauss or more. Such type of exotic systems with one or more statistical parameters in the above mentioned range are found in astrophysical bodies or in the very early universe. The commonly known particles are expected to show relatively different behavior in such exotic systems. Here we have to evaluate renormalization constants for different ranges of temperature and chemical potential, using corresponding limits of integration. These QED parameters are evaluated for all relevant ranges of temperatures and densities of interest in astrophysics and cosmology. Without getting in to the details of the FTD calculations, we can use the already existing expressions for these parameters and evaluate them for the relevant ranges. Statistical contribution to electron mass, charge, and wavefunction has been evaluated [12-18] for all different ranges of QED systems at extremely high temperatures and densities.

Light waves are transverse electromagnetic waves and move in free space with a constant speed c equal to 2.99×10⁸ m/s. It does not interact with the medium at low temperature and density and propagates through such media as if they travel through the vacuum. The dispersion of light in regular laboratory materials is supported by the transverse nature of light. When electromagnetic waves propagate through a medium, its dispersion in the medium is related to the distribution of matter in space. This matter can exist in different phases. However, at the temperatures of our interest in astrophysics and cosmology, matter is not in bound states of atoms and molecules. High-energy matter exists in the fluid form and high-energy particle physics approach is used to describe the dynamics of such materials. All the fluids with charged particles are either non-interacting ideal gases or they interact electromagnetically.

We are only considering the high-energy fluids and will not use kinetic theory of gases. These fluids are highly energetic fluids with all relativistically moving particles with variable densities. Different fluids behave differently based on its density. The inter-particle interaction of such relativistic fluids is treated as gauge interaction and follows the Feynman rules of QED gauge interaction, indicated in the given QED lagragian in the beginning. QED is a local theory so the scale of field theoretic study is the scale of quantum field theories. It allows the coexistence of variable phases of fluid in astrophysical systems, locally. The information about the correct phase can be found in terms of the parameters of the system.

i. **Non-interacting Ideal Gas**

A relativistic fluid of electrons and photons can exist in multiple forms at relativistic energies of electrons and extremely high temperatures of billions of Kelvin. A fluid at high energy with relativistically moving particles at extremely high energies is considered to be a fermigas of electrons because it follows Fermi spin statistics and does not interact at high energy.

Non-interacting relativistic fluid of electrons at extremely high temperature is considered as relativistic QED fluid. The energy is radiated in the form of electromagnetic waves here. An isolated heatbath is a closed system itself and not connected to any external system so the exchange of fluid or energy is not allowed. Universe and stellar cores are modeled as closed systems where everything satisfy the condition of thermodynamic equilibrium. This system can have one or more type of fluids but they all satisfy the equilibrium conditions throughout. Photon as quanta of electromagnetic radiation obey the Bose-Einstein distribution, given by Eqs. (2). Since the boson statistics allows an infinite number of particles in a state and we cannot put any limit on the number density of photons.

Non-interacting gas of electrons is the fermi-gas, which follow the spin statistics in a fluid at lowest energy and still is in thermal equilibrium with the rest of the contents of the heat bath. The fermigas of electrons follow spin statistics and are distributed using Fermi-Dirac distribution give in Eq. (3). Due to the Fermi statistics, no two electrons occupy exactly the same state. Only two particles can occupy the same state as a pair of spin-up and spin-down electrons and the occupation number comes out to be 2 for fermions. The number density of bosons is not limited but the number density of electrons is calculated using the spin statistics by integrating Eq. (3a) over all the possible energies. In the limit E > μ, the number density comes out to be

$$n_e = \sum_n (-1)^{n+1} \frac{T}{n} e^{-n\mu\beta} \qquad (5)$$

In the limit of small ratio of $\mu\beta = \frac{\mu}{T} \ll 1$, the exponential term quickly approaches to unity whereas for large values of n, the density will approach to zero exponentially. So the number density of electrons is related to the chemical potential directly. Larger number of electrons are related to smaller value of chemical potential and favors a non-interacting system.

In the limit of $\mu\beta = \frac{\mu}{T} > 1$, we integrate the distribution function in the negative energy with the upper limit $\mu = E$. The detailed analysis of the distribution makes it clear that we cannot have a bound positron system. The limit of energy $\mu = E$ corresponds to the Fermi energy and this fluid correspond to fermigas of electrons that fill up Landau levels and obey fermion statistics.

**ii.  Interacting Gas**

Interacting gas with low number density and ignorable chemical potential is still treated as an ideal gas, approximately. Bound states of electrons need the presence of positively charged nuclei and all the requirements of quantum mechanical description of atomic structure are still needed to be satisfied. These fermions are relativistically distributed in Landau levels in the presence of magnetic field created by the presence of charged fermions.

Free electrons can interact with ions in the fluid without bonding. This fluid can satisfy the conditions of plasma if the density and other conditions for the existence of plasma are satisfied. A quasi-neutral partially ionized gas is identified as a plasma when a fluid of charged and neutral particles is dense enough to show the collective behavior that is different from the individual behavior. Classical plasma fluid obeys the Maxwellian distribution and the average energy of the gas is proportional to its average temperature and individual mass of gas molecules. This temperature is related to the collision frequency of each type of molecules. Each fluid in a plasma may have a different kinetic energy and different temperature. In a classical electron-ion plasma, electrons interact with laser light. The laser accelerated high-energy plasma may generate large amplitude wakefields. When a photon propagates through a plasma, it interacts with the plasma waves and describe the excitation of wakefield [19-20] associated with laser-plasma acceleration through fast ignition fusion.

In classical plasma, number density plays an important role to determine the collision parameters including collision frequency and the mean free path between successive collisions. However, all the collision parameters in classical plasma are calculated using the Maxwellian distribution and bulk properties are determined by averaging over the Maxwellian distribution of velocities. The continuity equation of such plasmas is then written as:

$$\frac{\partial n}{\partial t} + n_0 \nabla \cdot v_e = 0 \qquad (6)$$

Now the rate of change of number density will depend on the rate of flow of the liquid associated with the change in composition of star and the rate of cooling or heating of star. On the other hand, propagation speed of particles is affected by the temperature and chemical potential. Therefore, this continuity equation will not work to give the correct results. Therefore, we need to look in to the conditions of QED plasma, for such a system and we discuss it in a little detail in the next section, using renormalization scheme of QED instead of the effective field theory to understand the particle behavior in plasma, in detail. The interaction of electron and photon is then calculated, using the Fermi-Dirac distribution instead of Maxwellian distribution, at high energies. This modification automatically incorporates the spin effect at high energies.

**iii.  QED Plasma**

QED plasma describes a relativistic system treated by the QED renormalization scheme at finite temperature and density (FTD). This covariant real-time formalism is related to the local conditions of the system and individual particle interactions with the medium. At extremely high energies, relativistic plasma of QED and physically measureable renormalized parameters of the theory are used to calculate the parameters of QED plasma [8-10].

Renormalization theory of QED implies that the longitudinal component of the photon is not zero, which allows the self-interaction of photons. Therefore, photon-photon scattering appears as a higher order effect in QED plasma and we discuss the impact of this longitudinal component on the electromagnetic properties of the medium. Photon-photon scattering crossection has not been measured yet but its indirect impact on laser-plasma interaction may be testable in the strong magnetic field. The strong magnetic field is available in stellar cores like neutron stars and the interaction of monochromatic signals with plasma may give much more details about the stellar cores, especially the superdense cores of neutron stars in general and magnetars, in particular. The vacuum birefringence is another effect. This will have an impact on the neutrino propagation, in the superdense media, due to the interaction of massive neutrinos with the magnetic field. However, this discussion is out of scope of this paper.

QED plasma provides a dispersive medium of weak and rapidly oscillating fields and nonlinear effects is expected due to the interaction with slowly varying strong fields. The strength of the magnetic field does not have a direct impact on QED parameters of magnetized plasmas. It may even lead to quantum magnetic collapse, as discussed in Ref. [5]. The high densities and associated high magnetic fields in QED plasma can describe the detailed structure of stars as well as indicate the presence of different phases in different regions of stellar cores.

When the temperatures and densities touch the exotic limits of superdense stars and other temporarily created experimental situations, we incorporate the spin statistics, use Fermi-Dirac and Bose-Einstein distributions, and deal with small regions of the medium instead of talking about the overall properties of the media as bulk properties. Bulk properties of such highly dynamic superdense bodies deal with averages that are determined by averages over the entire material. SO the drastically bulk properties cannot describe the detailed dynamics of the material and local behavior is drastically approximated. QED plasma is generated locally, but it has the ability to block, rather trap some radiation frequencies and had some of the information about the material and structure of the interior of the stellar bodies.

### iv.   Hard Thermal Loops and Superfluids

When the temperatures or energies are high enough to create heavy particles, QED is not the complete theory. An entire series of gauge theories including QED, electroweak and Quantum Chromodynamics (QCD) should be incorporated to describe such systems. At this point, discussion of particle interaction models is started and several parallel approaches for the extensions of the standard model, including supersymmetric models can be used. Since thermal field theories deal with the real particle background so the model dependent study with the particles, which are not yet observed, does not look like a very attractive approach. Therefore, the study of such behavior is done using the effective potential approach, which can incorporate the hard thermal loops that appears in the renormalization scheme when temperatures are so large. Thermal interactions with larger than electron mass are not ignorable any more. These systems are very complex and need to incorporate (so-called) hard-thermal loops [21] in relativistic fluids. Effective potential gives a workable approach for such systems. However, at such large temperatures and even higher chemical potentials, the phase of the matter is not known clearly. It may exist in the form of superfluids or quark-gluon plasmas, as well. We will not discuss this topic in detail because the calculational approach is out of scope of this paper and analytical results are not yet available due to the complexity of such systems. However, it can be easily noticed that all our calculations put a natural limit on temperatures and energies to make it a physically existing system in the form of fluid or QED plasma. Untreatable singularities provide the limit showing that failure to remove singularities mean the absence of such states in the physical world. As higher energies may produce heavier particles and their interaction is not necessarily QED interaction only, they contribute to some irremovable singularities at higher loop levels in QED and can be identified as hard loops. We are not dealing with the electroweakly interacting particles, QCD, or strong interactions in this paper so the discussion of the standard model is postponed as well.

Propagating particles interact with the medium through the virtually emitted particles that perturb the medium. These perturbative corrections grow in the medium and add mass and charge to the propagating particles. Temperature and chemical potential of electrons and the high-energy radiations in the medium can add up to some unphysical singularities. Renormalization scheme of QED provides a way to cancel singularities. Order by order cancellation of singularities is required for a physical theory due to Kinoshita-Lee and Neunberg (KLN-) theorem [22]. This order-by-order cancelation of singularities makes it possible to calculate the renormalization constant of QED as physically measurable parameters of the theory at each level of perturbation and correspond to electron mass, wavefunction and charge. Since the QED coupling is proportional to the square of electron charge, and therefore, has a measureable effect on the electromagnetic properties of the medium itself. Therefore, electromagnetic signal carries the information about the electromagnetic properties of the medium, which may include the information on the phase of matter along with the composition of stars.

Statistical corrections to the renormalization constants will be discussed in detail in the next section and their relevance to the measureable parameters of astrophysics will be discussed there as well. However, description of the connection of wave propagation in an interacting medium needs a much better physical understanding to update the measurement techniques.

3. **Propagation of Light**

   The information about the correct phase can be found in terms of the parameters of the system. The masslessness of photon requires $K^2 = 0$, which leads to the vanishing of longitudinal component of electromagnetic waves and transeverasality of em-signals.

   The spherical nature of light is associated with the fact that electromagnetic waves satisfy the equation of a sphere, using electric field **E**, magnetic field **B** and the propagation vector **k**. For such a system $\mathbf{E}^2 + \mathbf{B}^2 + \mathbf{k}^2 = \omega^2$, for constant value of the angular velocity $\omega$, such that **E, B and k** vectors are all perpendicular to one another in natural units. This equation is always satisfied in 4-dimensional space and requires the massless photon as a gauge requirement. This puts a constraint on self-interaction of photon leading to the transverse nature of light and does not allow electromagnetic waves to interact with the medium.

   A detailed mechanism of propagation in a three dimensional space satisfies the conditions of spherical growth of an em-beam in the medium that depends on the three dimensional components of the wavenumber of light waves. The propagation of longitudinal and transverse components in space is associated with the angular frequencies as the fourth component and can change the energy because the massless of energy quanta cannot be assumed in an interacting medium. The four dimensional dispersion of electromagnetic waves with nonzero longitudinal component is generally expressed as:

$$\pi_{\mu\nu} = \begin{pmatrix} -\frac{k^2}{K^2}\pi_L & -\frac{i\omega k_1}{K^2}\pi_L & -\frac{i\omega k_2}{K^2}\pi_L & -\frac{i\omega k_3}{K^2}\pi_L \\ -\frac{i\omega k_1}{K^2}\pi_L & \left(-1-\frac{k_1^2}{k^2}\right)\pi_T + \left(\frac{\omega^2 k_1^2}{k^2 K^2}\right)\pi_L & \left(-\frac{k_1 k_2}{k^2}\right)\pi_T + \left(\frac{\omega^2 k_1 k_2}{k^2 K^2}\right)\pi_L & \left(-\frac{k_1 k_3}{k^2}\right)\pi_T + \left(\frac{\omega^2 k_1 k_3}{k^2 K^2}\right)\pi_L \\ -\frac{i\omega k_2}{K^2}\pi_L & \left(-\frac{k_1 k_2}{k^2}\right)\pi_T + \left(\frac{\omega^2 k_1 k_2}{k^2 K^2}\right)\pi_L & \left(-1-\frac{k_2^2}{k^2}\right)\pi_T + \left(\frac{\omega^2 k_2^2}{k^2 K^2}\right)\pi_L & \left(-\frac{k_2 k_3}{k^2}\right)\pi_T + \left(\frac{\omega^2 k_2 k_3}{k^2 K^2}\right)\pi_L \\ -\frac{i\omega k_3}{K^2}\pi_L & \left(-\frac{k_1 k_3}{k^2}\right)\pi_T + \left(\frac{\omega^2 k_1 k_3}{k^2 K^2}\right)\pi_L & \left(-\frac{k_2 k_3}{k^2}\right)\pi_T + \left(\frac{\omega^2 k_2 k_3}{k^2 K^2}\right)\pi_L & \left(-1-\frac{k_3^2}{k^2}\right)\pi_T + \left(\frac{\omega^2 k_3^2}{k^2 K^2}\right)\pi_L \end{pmatrix}$$

Whereas, the polarization tensor $\pi_{\mu\nu}(K)$ depends on the 4-dimensional K and is given as:

$$K^2 = \omega^2 - k^2, \qquad \omega = K_\alpha u^\alpha.$$

and is related to each component of three momentum and energy of photon as

$$(p_1, p_2, p_3; E) = \hbar(k_1, k_2, k_3, \omega)$$

The longitudinal and transverse components of the vacuum polarization tensor are generally represented as $\pi_L(k,\omega)$ and $\pi_T(k,\omega)$, respectively. The vacuum polarization tensor is rewritten as a linear combination of longitudinal and transversely polarized tensor $Q_{\mu\nu}$ and $P_{\mu\nu}$ respectively as:

$$\pi_{\mu\nu}(K,\mu) = P_{\mu\nu}\pi_T(K,\mu) + Q_{\mu\nu}\pi_L(K,\mu)$$

Such that

$$P_{\mu\nu} = \tilde{g}_{\mu\nu} + \frac{\widetilde{K}_\mu \widetilde{K}_\nu}{k^2}$$

and

$$Q_{\mu\nu} = -\frac{1}{K^2 k^2}\left(k^2 u_\mu + \omega \widetilde{K}_\mu\right)\left(k^2 u_\nu + \omega \widetilde{K}_\nu\right),$$

in the usual notation of QED, whereas, the 4-velocity of the heatbath is $u_\mu$ and written in the covariant form as $u_\mu = (1,0,0,0)$: a static heat bath with a normalized energy change, following the conditions of modified covariant tensors as:

$$\tilde{g}_{\mu\nu} = g_{\mu\nu} - u_\mu u_\nu$$

and

$$\widetilde{K}_\mu = K_\mu - \omega u_\mu$$

Such that they satisfy the conditions:
From the above relations, we can write the complete dispersion of longitudinal component $P_{\mu\nu}$ as:

$$P_{\mu\nu} = \begin{pmatrix} 0 & 0 & 0 & 0 \\ 0 & -1 - \frac{k_1^2}{k^2} & -\frac{k_1 k_2}{k^2} & -\frac{k_1 k_3}{k^2} \\ 0 & -\frac{k_1 k_2}{k^2} & -1 - \frac{k_2^2}{k^2} & -\frac{k_2 k_3}{k^2} \\ 0 & -\frac{k_1 k_3}{k^2} & -\frac{k_2 k_3}{k^2} & -1 - \frac{k_3^2}{k^2} \end{pmatrix} \qquad (5a)$$

and the transverse components $Q_{\mu\nu}$

$$Q_{\mu\nu} = \begin{pmatrix} -\dfrac{k^2}{K^2} & -\dfrac{i\omega k_1}{K^2} & -\dfrac{i\omega k_2}{K^2} & -\dfrac{i\omega k_3}{K^2} \\ -\dfrac{i\omega k_1}{K^2} & \dfrac{\omega^2 k_1^2}{k^2 K^2} & \dfrac{\omega^2 k_1 k_2}{k^2 K^2} & \dfrac{\omega^2 k_1 k_3}{k^2 K^2} \\ -\dfrac{i\omega k_2}{K^2} & \dfrac{\omega^2 k_1 k_2}{k^2 K^2} & \dfrac{\omega^2 k_2^2}{k^2 K^2} & \dfrac{\omega^2 k_2 k_3}{k^2 K^2} \\ -\dfrac{i\omega k_3}{K^2} & \dfrac{\omega^2 k_1 k_3}{k^2 K^2} & \dfrac{\omega^2 k_2 k_3}{k^2 K^2} & \dfrac{\omega^2 k_3^2}{k^2 K^2} \end{pmatrix} \quad (5b)$$

The dispersion of signals is subject to the signal energy and wavelength, whereas, the statistical parameter dependence comes from the calculation of the components of polarization tensor, and is discussed in the next section.

4. **QED Parameters in different ranges of temperature and chemical potential**

It has been previously calculated that the electron mass, wavefunction and charge can grow with temperature and chemical potential while propagating through the extremely hot and dense backgrounds of astronomical systems. Using the renormalization scheme of QED, the background contribution of statistical variables to the measurable quantities such as the electron mass, wavefunction and charge is calculated in literature, using different approaches such as the renormalization scheme or the effective potential approach. All of these QED parameters become a function of temperature, density and magnetic field due to its interaction with the relevant medium. The dependence on the magnetic field is not relevant for QED processes unless we talk about exceptionally magnetized stars such as magnetars only but it is ignored for the current study. Temperature and chemical potential dependence is incorporated through the radiative corrections using the renormalization scheme of QED in real-time formalism [6-18] in Minkowski space.

The statistical contribution to self-energy correction comes out to be:

$$\frac{\delta m}{m}(T,\mu) \approx \frac{\alpha \pi T^2}{3m^2}\left[1 - \frac{6}{\pi^2}c(m\beta,\mu)\right] + \frac{2\alpha}{\pi}\frac{T}{m}a(m\beta,\mu) - \frac{3\alpha}{\pi}b(m\beta,\mu). \quad (7)$$

It leads to the FTD contribution to the renormalization constant of the wavefunction of electron as:

$$Z_2^{-1}(T,\mu) = Z_2^{-1}(T=0,\mu) - \frac{2\alpha}{\pi}\int \frac{dk}{k}n_B(k) - \frac{5\alpha}{\pi}b(m\beta,\mu) + \frac{\alpha}{\pi}\frac{T^2}{E^2}\frac{1}{v}\ln\frac{1-v}{1+v} \times$$

$$\left[c(m\beta,\mu) - \frac{\pi^2}{6} - \frac{m}{T}a(m\beta,\mu)\right] \quad (8)$$

Whereas, the charge renormalization constant can be written as:

$$Z_3(T,\mu) \approx 1 + \frac{\alpha T^2}{6m^2}\left[\frac{ma(m\beta,\mu)}{T} - c(m\beta,\mu) + \frac{1}{4}(\ m^2 \right.$$
$$\left. + \frac{\omega^2}{3}b(m\beta,\mu)\right] \qquad (9)$$

These two constants are evaluated in terms of ABC functions introduced by Masood,et.al in literature. These functions appear in all type of perturbative corrections due to the integration of Fermi-Dirac function with different values of energy. A detailed description of these functions in different limits of temperatures and densities is given in the appendix of Ref.[24]. In the high-temperature limit $(T > m > \mu)$, these functions appear as [12-13]:

$$a(m\beta,\pm\mu) = \ln(1 + e^{-(m\pm\mu)\beta})$$

$$b(m\beta,\pm\mu) = \sum_{n=1}^{\infty}(-1)^n e^{\mp n\beta\mu} Ei(-nm\beta)$$
$$c(m\beta,\pm\mu) = \sum_{n=1}^{\infty}\frac{(-1)^n}{n^2} e^{-n(m\pm\mu)\beta}$$
$$d(m\beta,\pm\mu) = \sum_{n=1}^{\infty}\frac{(-1)^n}{n^3} e^{-n(m\pm\mu)\beta}$$

Here, $Ei(x)$ represents an exponential integral, given as:
$$Ei(x) = \int_{-x}^{\infty}\frac{e^{-t}dt}{t}$$

In the high chemical-potential limit $(T < m < \mu)$, we obtain [15]:

$$a(m\beta,-\mu) = \mu - m - \sum_{l=0}^{\infty}\frac{(-1)^l}{\mu^l}\sum_{n=1}^{\infty}\frac{(-1)^n}{(n\beta)^{1-l}} e^{-n\beta(m-\mu)}$$

$$b(m\beta,-\mu) = \ln(\mu/m)$$

$$c(m\beta,-\mu) = \frac{\mu^2 - m^2}{2} - \sum_{l=0}^{\infty}\frac{(-1)^l}{\mu^l}\sum_{n=1}^{\infty}\frac{(-1)^n}{(n\beta)^l} e^{-n\beta(m-\mu)}$$

$$d(m\beta,-\mu) = \frac{\mu^3 - m^3}{3} - \sum_{l=0}^{\infty}\frac{(-1)^l}{\mu^l}\sum_{n=1}^{\infty}\frac{(-1)^n}{(n\beta)^{l+1}} e^{-n\beta(m-\mu)}$$

These Masood functions evaluate the background fermion contributions and start to appear at the first order in alpha in perturbative study. At low temperatures and low densities, they simply vanish because of the absence and very low concentration of fermions. In other words, it measures the contribution of interaction with matter. Computation of these functions for extreme cases correspond to the early universe or the superdense cores of stars and these results are discussed later.

### 4.1 Lamb Shift

The lamb shift is a purely measureable perturbative effect, which causes a small change in wavelength. Knight [25] has looked at the thermal effects on the Lamb shift. A small shift in the bound state energies is calculated from the regular quantum mechanical energy values corresponding to the integra values on n such that:

$$E_n = -\left(\frac{me^4}{8(nh\varepsilon_0)^2}\right) \qquad (10\text{ a})$$

$$E = E_f - E_i = \left(\frac{me^4}{8(nh\varepsilon_0)^2}\right)\left(\frac{1}{n_i^2} - \frac{1}{n_f^2}\right) \qquad (10\text{ b})$$

The change in the electron mass will affect the energies of the energy levels. Comparing these formulae with the Rydberg formula gives a shift on energy between successive energy levels as

$$R_H = -\left(\frac{me^4}{8h^3\varepsilon_0^2}\right) \qquad (11)$$

and, mass of electrons changes with temperature and chemical potential in different FTD conditions and are given in Eq. (5) and (6). Therefore, the chemical potential and temperature dependence is simply introduced by the electron mass.

On the other hand the total energy shift due to vacuum polarization is determined as :

$$\Delta E = \int d^3r \, \Delta V(r)|\psi(r)|^2$$

where

$$\Delta V(r) = \frac{e^2}{(2\pi)^3} \int d^3q \, e^{iq\cdot r} \left[\frac{\widetilde{\Pi}(q)}{q^2}\right]$$

And

$$\Delta E_n = \frac{-4\alpha^5 m}{15\pi n^3} \qquad (12)$$

n is the Landau number in Eq. (11) and is associated with the Landau levels of energy for the Fermigas of electron. This correction may not be relevant for atoms for temperature dependent electron mass as the existence of atoms is not expected at $10^{10}$ K. However, high chemical potential is not necessarily ignorable for large molecules.

### 5. Propagation of electromagnetic waves in a medium

The propagating light through a quantum statistical medium interacts with the medium and picks up FTD corrections from the medium as its charge can be renormalized at high temperatures. These photons inside the medium are identified as plasmons as they are the quanta of energy waves propagating inside the medium. The longitudinal and transverse components of plasmons depend on statistical parameters of the medium but this effect depends on the frequency (energy) and wavenumber (momentum) of the electromagnetic waves because now they are not transverse or spherical waves. This change has a great impact on the electromagnetic properties of the medium due to the background fermion (like electrons) contribution and the gauge coupling (QED coupling), both of these constants of QED theory become temperature and chemical potential dependent parameters. This medium can be in a state of Fermigas or QED plasma, depending on the statistical conditions. When the electromagnetic waves are emitted after travelling through such a system and can be used as a powerful source of information. These waves are observed and analyzed as a carrier of the properties of the medium and these properties are determined from the behavior of these signals. The longitudinal and transverse components of electromagnetic waves have been calculated as:

$$\pi_L \cong \frac{4e^2}{\pi^2}\left(1 - \frac{\omega^2}{k^2}\right)\left[\left(1 - \frac{\omega}{2k}\ln\frac{\omega+k}{\omega-k}\right)\left(\frac{ma(m\beta,\mu)}{\beta} - \frac{c(m\beta,\mu)}{\beta^2}\right) + \frac{1}{4}\left(2m^2 - \omega^2 + \frac{11k^2+37\omega^2}{72}\right)b(m\beta,\mu)\right] \quad (13a)$$

$$\pi_T \cong \frac{2e^2}{\pi^2}\left[\left\{\frac{\omega^2}{k^2} + \left(1 - \frac{\omega^2}{k^2}\right)\ln\frac{\omega+k}{\omega-k}\right\}\left(\frac{ma(m\beta,\mu)}{\beta} - \frac{c(m\beta,\mu)}{\beta^2}\right) \right.$$
$$\left. + \frac{1}{8}\left(2m^2 + \omega^2 + \frac{107\omega^2 - 131k^2}{72}\right)b(m\beta,\mu)\right] \quad (13b)$$

$$\varepsilon(k) = 1 - \frac{\pi_L(K)}{k^2} \qquad \frac{1}{\mu(k)} = \frac{k^2\pi_T(K) - \omega^2\pi_L(K)}{k^2 K^2} \quad (14a)$$

$$v_{prop} = \sqrt{\frac{1}{\varepsilon(K)\mu(K)}} \qquad r_i = \frac{c}{v} = \sqrt{\frac{\varepsilon(K)\mu(K)}{\varepsilon_0(K)\mu_0(K)}} \quad (14b)$$

We can substitute the values of $\pi_T(K,T,\mu)$ and $\pi_L(K,T,\mu)$ from Eqn. (11) to find all the temperature and chemical potential dependent values of propagation velocity, refractive index, electric permittivity and magnetic permeability.

The calculation of the longitudinal and transverse components of the photon frequency $\omega$ and the momentum k at different temperatures for given values of $\omega$ and k lead to find the actual measureable wavenumber and frequency using the relations:

$$k = (k_L^2 + k_T^2)^{1/2} \quad (15\ a)$$

Such that

$$\omega = (\omega_L^2 + \omega_T^2)^{1/2} \quad (15\ b)$$

However, their effective values are determined from the corresponding values of polarization tensors for the relevant ranges of temperature and chemical potential.

The longitudinal component of the vacuum polarization tensor at finite temperature can be used to determine the phase of the medium indicating overall properties of the background. Medium properties are changing with temperature due to modified electromagnetic couplings (in Eq. 2c). $K_L$ is evaluated by taking $\omega = k_0 = 0$ and p very small, the Debye shielding length $\lambda_D$ of such a medium is then given by the inverse of $K_L$

$$\lambda_D = 1/K_L \tag{16 a}$$

and the corresponding $\omega_D$ is

$$\omega_D = \frac{2\pi v_{prop}}{\lambda_D} = 2\pi K_L v_{prop} \tag{16 b}$$

Satisfying the relation

$$f_D \lambda_D = v_D \tag{16 c}$$

The plasma frequency $\omega_P$ is also related to $\omega_T$ from $\Pi_T(\omega = |k|)$ as

$$\omega_D = \frac{2\pi v_D}{\lambda_D} \tag{17}$$

and it is defined as $\omega_P{}^2 = \omega_T{}^2$ as $\omega_L{}^2 = 0$, and Debye shielding length is obtained from the longitudinal component of wavelength using equation (10 b). These results can easily be generalized to different situations using the initial values of $\pi_L$ and $\pi_T$.

6. **Results and Discussions**

We evaluate all the quantities for the early universe (high temperature T) and for superdense materials (large chemical potential $\mu$) in the extreme cases. All the limits of integrations can be determined from the scope of the equipment or physical conditions. For extremely large temperatures of the early universe and the ignorable densities, the only contributing term is proportional to the ratio of temperature and the electron mass,

$$\frac{\delta m(T)}{m} \approx \frac{\alpha \pi T^2}{2m^2} \tag{18 a}$$

and the renormalized mass, which is a physically measureable mass,

$$m_R = m + \delta m(T=0) + \delta m(T) \tag{18 b}$$

In the above equations, the only contributing factor due to the fermion interaction, gives:

$$c(m\beta, \mu) \approx -\frac{\pi^2}{12}$$

Since E is of the order of T for extremely large temperatures and we can simplify the log-term as

$$\frac{1}{v} ln \frac{1-v}{1+v} \approx -2 \quad \text{(for E} \gg \text{m)}$$

The net result for the wavefunction of electron is then given as:

$$Z_2^{-1}(T,\mu) \approx \frac{2\alpha}{\pi} \int \frac{dk}{k} n_B(k) + \frac{\alpha\pi T^2}{6m^2} \quad (19\ a)$$

$$\frac{2}{\pi} \int \frac{dk}{k} n_B(k) = \varepsilon(k) \quad (19\ b)$$

The new small parameter $\varepsilon(k)$ is introduced as it gives a constant small value that varies with k slowly, with dominant logarithmic behavior. At high temperatures, the lower limit of k is T and the upper limit is set by the physically measureable value of k. Therefore, the statistically renormalized wavefunction $\Psi_R(p,T)$ corresponds to the physically measureable form of the wavefunction in terms of related probability of finding electron at the given energy (T) as:

$$\Psi_R(p,T) = [Z_2^{-1}(T=0) + Z_2^{-1}(T,\mu)]\Psi \quad (20)$$

And the electric charge `e' is renormalized for a given temperature $e_R(p,T)$ and is written as

$$e_R(p,T) = [Z_3^{1/2}(T=0) + Z_3^{1/2}(T,\mu)]e, \quad (21\ a)$$

giving the corresponding renormalized QED coupling parameter $\alpha_R$

$$\alpha_R = [Z_2^{-1}(T=0) + Z_2^{-1}(T,\mu)]\alpha \quad (21\ b)$$

However, the QED coupling is expressed as:

$$\alpha(T) = \alpha(T=0)\left(1 + \frac{\alpha^2 T^2}{6m^2}\right) \quad T \gg m \quad (21\ c)$$

and the number density of electrons still satisfy the fermi rule for spin ½ particles as:

$$n_e = T \sum_n \frac{(-1)^{n+1}}{n} \approx \frac{T}{2} \quad (22)$$

A detailed discussion about the electromagnetic properties of the medium is included in Ref. [5-8]. It shows that at high temperatures, due to the selfmass of photon, the medium can behave as a plasma, even for short period. QED plasma parameters such as the Debye shielding and the plasma frequencies are calculated at FTD, as well.

At high densities, in superdense media, the systems behave almost like a classical system as the parameter $\beta\mu$ may have values comparable to a classical fluid even for extremely large values of $\mu$ and T. For such systems, the chemical potential corresponds to the lowest energy (Fermi energy) of the fermigas of electrons. We present thermal contributions to the renormalization constants of QED for ignorable chemical potential values in Table 1 to give some idea of comparison of contribution for different parameters. However, the calculation of actual contributions, the given set of statistical parameters is calculated in a straightforward way from the provided set of equations. These approximations are relevant for the early universe.

Table 1: Thermal contribution to QED renormalization constants at T > m for small µ

| Temp(T/m) | Selfmass | charge | coupling | wavefunction |
|---|---|---|---|---|
| 1 | 0.022941 | 0.002433 | 0.001217 | 0.003823 |
| 1.25 | 0.035845 | 0.003802 | 0.001901 | 0.005974 |
| 1.5 | 0.051616 | 0.005474 | 0.002737 | 0.008603 |
| 1.75 | 0.070255 | 0.007451 | 0.003726 | 0.011709 |
| 2 | 0.091762 | 0.009732 | 0.004866 | 0.015294 |
| 2.25 | 0.116137 | 0.012318 | 0.006159 | 0.019356 |
| 2.5 | 0.143379 | 0.015207 | 0.007603 | 0.023896 |
| 2.75 | 0.173488 | 0.0184 | 0.0092 | 0.028915 |
| 3 | 0.206465 | 0.021898 | 0.010949 | 0.034411 |
| 3.25 | 0.24231 | 0.0257 | 0.01285 | 0.040385 |
| 3.5 | 0.281022 | 0.029805 | 0.014903 | 0.046837 |
| 3.75 | 0.322602 | 0.034215 | 0.017108 | 0.053767 |
| 4 | 0.367049 | 0.038929 | 0.019465 | 0.061175 |
| 4.25 | 0.414364 | 0.043948 | 0.021974 | 0.069061 |
| 4.5 | 0.464546 | 0.04927 | 0.024635 | 0.077424 |
| 4.75 | 0.517596 | 0.054897 | 0.027448 | 0.086266 |
| 5 | 0.573514 | 0.060827 | 0.030414 | 0.095586 |

Table 1 shows that the thermal corrections are adding mass relatively quickly. Even at the temperature below 2MeV (around nucleosynthesis in the early universe) can add 45% mass to electron, whereas thermal contribution to the coupling is just 3% at these temperatures.

In superdense systems ( m < T < $\mu$), the QED type perturbative contributions at the largest possible known ranges of chemical potential has already been evaluated for relevant ranges of photon energies and the statistical conditions using the given form of Masood's functions, as discussed in detail in Ref. [6]. The selfmass of electron in this case is

$$\frac{\delta m(T)}{m} \approx -\frac{\alpha}{\pi}\left[3 ln\frac{\mu}{m} + 4 - \left(\frac{\mu}{m}\right)^2\right] \quad (23\ a)$$

$$e = -\frac{\alpha}{4}\left(\frac{\mu}{m}\right)^2 \quad (23\ b)$$

Whereas thermal contributions to the wavefunction renormalization constant is not affected significantly. Density dependent contribution will reduce the mass for low densities, which is not physically possible so the physical systems with $\beta\mu < 3$ are not allowed to exist. Also, the QED coupling is decreased with the increase in density.

Table 2: Density contribution to QED renormalization constants at $\mu$ >m for small

| $\mu$/m | mass | mass | charge | coupling |
|---|---|---|---|---|
| 1 | 0 | -0.00697 | -0.00182 | -0.00091 |
| 1.25 | -0.00155 | -0.00722 | -0.00285 | -0.00143 |
| 1.5 | -0.00283 | -0.00689 | -0.00411 | -0.00205 |
| 1.75 | -0.0039 | -0.00608 | -0.00559 | -0.00279 |
| 2 | -0.00483 | -0.00483 | -0.0073 | -0.00365 |
| 2.25 | -0.00565 | -0.00318 | -0.00924 | -0.00462 |
| 2.5 | -0.00638 | -0.00116 | -0.01141 | -0.0057 |
| 2.75 | -0.00705 | 0.001226 | -0.0138 | -0.0069 |
| 3 | -0.00765 | 0.003958 | -0.01642 | -0.00821 |
| 3.25 | -0.00821 | 0.007029 | -0.01927 | -0.00964 |
| 3.5 | -0.00873 | 0.010432 | -0.02235 | -0.01118 |
| 3.75 | -0.00921 | 0.014161 | -0.02566 | -0.01283 |
| 4 | -0.00966 | 0.018211 | -0.0292 | -0.0146 |
| 4.25 | -0.01008 | 0.022579 | -0.03296 | -0.01648 |
| 4.5 | -0.01048 | 0.027261 | -0.03695 | -0.01848 |
| 4.75 | -0.01086 | 0.032255 | -0.04117 | -0.02059 |
| 5 | -0.01121 | 0.037559 | -0.04562 | -0.02281 |
| 5.25 | -0.01155 | 0.04317 | -0.0503 | -0.02515 |
| 5.5 | -0.01188 | 0.049088 | -0.0552 | -0.0276 |
| 5.75 | -0.01219 | 0.05531 | -0.06033 | -0.03017 |
| 6 | -0.01248 | 0.061836 | -0.06569 | -0.03285 |
| 6.25 | -0.01277 | 0.068664 | -0.07128 | -0.03564 |

Values of renormalization constants can simply be used to determine the order of magnitude contribution to the decay rates, lamb shift and different processes, even without getting in to details of individual radiative contributions to each and every astrophysical process.

7. **Conclusions**

It can be seen in Eqns. (20-22), thermal contribution is always proportional to $\frac{T^2}{m^2}$ giving a hyperbolic dependence on temperature which changes to the exponential dependence with temperature because higher order in alpha terms start to contribute for higher energies.

However, all of these parameters are controlled by the chemical potential of the system less aggressively. At low values of $\frac{\mu}{T}$, it either depends on the first power of chemical potential in units of electron mass or its higher powers. Also, the Fermi statistics limit the number density of electrons managing its distribution and maximum value of chemical potential is achieved, quantum mechanically, as Fermi energy.

QED coupling, refractive index, plasma frequency, Debye shielding length and electromagnetic properties can all be calculated in terms of the QED parameters, using FTD contributions. It has also been noticed that certain values of the momenta and energy of Plasmon are trapped in the medium and may not be emitted due to bending back in to the medium.

Such systems may exist in several interchangeable states of matter locally depending on the local statistical properties of the interacting gas or fluid, which may exist temporarily in the form of relativistic QED plasma state under suitable conditions. State of matter and statistical parameters are closely related to each other. This astrophysical matter found in variable states like relativistic fluid, plasma, superfluid or even Fermi gas, based on its composition, temperature and density. The state of matter is determined by the properties of light in that medium. Existence of different phases at quantum scale indicates frequent phase transition at high temperatures and densities for rapidly rotating superdense material of neutron stars, due to high energy and high angular momenta. Information of phase indicates the range of statistical variables and composition that has a direct relation with the electromagnetic interaction properties of medium itself. We, therefore, recommend to incorporate FTD corrected parameters to analyze astronomical data to obtain more detailed information about the astronomical bodies.